%% file: turb.tex
\documentstyle[epsfig]{europhys}

\input euromacr

\begin{document}

\euro{X}{X}{1-6}{X}
\Date{X}
\shorttitle{E. Trizac~: A coalescence model for decaying 2D turbulence} 
\title{A coalescence model for freely decaying two-dimensional 
turbulence}
\author{Emmanuel TRIZAC~\footnote{Present address: 
FOM Institute for Atomic and Molecular Physics,
Kruislaan 407,  1098 SJ Amsterdam (The Netherlands). E-mail: 
etrizac@amolf.nl}
}
\institute{Laboratoire de Physique (URA 1325 du CNRS), 
Ecole normale sup\'erieure de Lyon \\
69364 Lyon Cedex 07 (France)}

\rec{X}{X}

\pacs{
\Pacs{47}{27.Eg}{Turbulence simulation and modeling}
\Pacs{05}{70.Ln}{Nonequilibrium thermodynamics, irreversible processes}
\Pacs{02}{70.Ns}{Molecular Dynamics and particle methods}
}

\maketitle

\begin{abstract}
We propose a ballistic coalescence model (punctuated-Hamiltonian approach)
mimicking the fusion 
of vortices in freely decaying two-dimensional turbulence.
A temporal scaling behaviour is reached where the vortex density 
evolves like $t^{-\xi}$. A mean-field analytical argument
yielding the approximation $\xi=4/5$ is shown to slightly overestimate
the decay exponent $\xi$ whereas Molecular Dynamics simulations 
give $\xi =0.71\pm 0.01$, in agreement with recent laboratory
experiments and simulations of Navier-Stokes equation.

\end{abstract}

Two-dimensional turbulence has the fascinating property
of organizing into coherent structures from a disordered 
background. This feature has been observed in laboratory
experiments \cite{CoBa,vHFl} and in numerical simulations,
as first emphasized and investigated by McWilliams
\cite{macWilliams}. It can account for the robustness
of Jupiter's Great Red Spot, a large scale vortex which has been
persisting in a turbulent shear for more than three centuries
\cite{Sommeria}.

We shall address the issue of freely decaying two-dimensional
turbulence, for which the decay from random initial conditions
can be divided into three stages. In the first stage, 
the system self-organizes into a set of coherent vortices
containing most of the flow vorticity. The evolution is then dominated
by the mutual advection of these structures, punctuated by 
dissipative events: whenever two like-sign vortices come 
closer than a critical distance, they merge (coalesce) to form a bigger 
vortex, while dipoles form a very stable state.
Finally, the third stage starts when there are 
very few dipoles left, which decay diffusively.

In this Letter, we shall concentrate on the second stage of
the evolution, during which the average extension 
of the vortices and their relative distance
grow whereas their number density $n$ decreases
in time. Numerical simulations of the Navier-Stokes equation
\cite{CmWPWY,Weiss93} have shown that this decay is algebraic:
$n(t) \propto t^{-\xi}$ with $\xi \simeq 0.72 \pm 0.03$.
On the experimental side, recent investigations in thin
stratified layers of electrolyte emphasize the importance of
coherent vortex dynamics for the decay of turbulence in 
two dimensions
\cite{Marteau,HansenTabeling}.

Assuming a self-similar evolution of the vortex system, one can
infer from dimensional grounds that $n(t) \propto t^{-2}$, provided
the energy is the only conserved quantity \cite{Weiss93,Batchelor}.
However, the emergence of coherent structures significantly 
slows the density decay ($\xi < 2$). In addition, numerical
solutions \cite{CmWPWY,Weiss93} indicate the appearance
of a second conserved quantity: the vorticity amplitude $\omega$
inside the vortex cores. On the assumption that 
all the vorticity is concentrated in the vortices and given the 
two above mentioned invariants, the authors of \cite{CmWPWY,Weiss93}
derived a non-conservative scaling theory expressing all
statistical properties in terms of the scaling exponent $\xi$, 
and constructed the simplest model capturing the essential features 
of the self-similar stage of freely decaying two-dimensional
turbulence. As no inviscid invariants other than the 
energy and the typical vorticity are preserved, 
their theory is non-conservative.
The vortices are modelized by discs of
radii $\sigma_i$, uniform vorticity $\omega$ and thus circulation
$\Gamma_i = \pi \,\sigma_i^2 \omega$. The equation of motion 
of vortex $i$, having position 
${\bf r}_i(t)=\{x_i(t),y_i(t)\}$, is assumed
to be governed by the Hamiltonian dynamics of point-vortices
(the so-called Kirchoff's laws \cite{Kirchoff}):
\begin{equation}
\Gamma_i  \left(\begin{array}{c}
   \dot x_i\\
   \dot y_i
\end{array}\right) = 
\left(\begin{array}{r}
\! \! -\, \partial_{y_i} {\cal H}   \\
\! \!  \partial_{x_i} {\cal H}
\end{array}\right), 
\qquad \dot\Gamma_i = 0
\label{eq:kir}
\end{equation}
where the Hamiltonian $\cal H$ takes the form
\begin{equation}
{\cal H} = \frac{1}{2}\,\sum_{i,j}\, \Gamma_i\Gamma_j\, 
G\left({\bf r}_i(t),{\bf r}_j(t)\right)
\end{equation}
and $G$ is the Green's function inverting the laplacian $\nabla^2$
with the required boundary conditions ({\it e.g.}
Dirichlet-like in a closed domain). The dynamics
(preserving all moments of the vorticity) is punctuated by
inelastic merging events with a  rule derived from the conservation laws:
since the energy $\cal E$ of a vortex is of order 
$\Gamma^2 \propto \omega^2\,\sigma^4$, the local conservation 
of $\cal E$ with a time independent vorticity implies that the vortex
resulting from the coalescence of vortices $i$ and $j$ 
has a radius $\sigma_{ij}$ given by 
\begin{equation}
\sigma_{ij} = (\sigma_i^4+\sigma_j^4)^{1/4}.
\label{eq:sigma4}
\end{equation}
This punctuated-Hamiltonian approach has numerically
given a decay exponent $\xi\simeq 0.72 \pm 0.02$
(cf \cite{CmWPWY,Weiss93}), in agreement with the
simulations of the full Navier-Stoke equation, indicating
that the restriction of vortex interactions to
those that have the most dramatic impact on the flow 
({\it i.e.} those in which the number of vortices
changes), seems sufficient to account for the decay exponent 
$\xi$. 

Several theoretical attempts have been made to determine
$\xi$. Motivated by the scaling laws of \cite{Weiss93},
Pomeau proposed an argument yielding $\xi=1$, but argued for
lowering corrections \cite{Pomeau}. In the possibly related context of
Ginzburg-Landau vortex turbulence, the value $\xi=3/4$ has been
put forward \cite{HuAl}. On the other hand, mapping the original problem 
on a solvable model of charges on a lattice, Sire obtained
$n(t) \propto \ln(t)/t$, where the logarithmic correction 
could explain why an effective exponent $\xi$ lower than 1
is measured \cite{sire}. So far, the accuracy associated with 
system sizes available for the simulations
or the experiments does not allow to discriminate such a law
against a truly algebraic behaviour. 

In this Letter, we shall further simplify the original 
punctuated-Hamiltonian model of 
Carnevale {\it et al.}, then obtain a simple
analytical estimation for $\xi$ and test its validity by
extensive Molecular Dynamics simulations. 
We modelize the motion
of vortices between merging events 
by free flights, and concentrate on the asymptotic scaling
regime followed by such a ballistic coalescence model,
which will be shown to capture the essential features of the
decay stage. A system 
initially made up of $N_0$ identical discs of radius
$\sigma_0$ in a two-dimensional domain of surface $S$, with
velocities ${\bf v}_i$ sampled from a distribution $\varphi_0({\bf v})$,
undergoes free particle motion, interrupted by a succession 
of instantaneous, completely inelastic collisions whenever a pair of
discs comes into contact. As a result of a binary collision,
the two initial discs merge into a single disc, which radius is given 
by eq (\ref{eq:sigma4}). The corresponding merging rule 
for the velocity is needed to completely specify the model, 
but we shall indicate that its precise form does not 
affect the asymptotic properties of the scaling regime.
A particular model of ballistic coalescence with conservation of
mass and momentum has been introduced in \cite{CPY}.
We first present a heuristic argument justifying
the occurrence of an algebraic asymptotic regime on dimensional
grounds.
From the density $n(t)=N(t)/S$, the 
instantaneous average radius $\langle \sigma(t)\rangle$, 
the instantaneous average modulus of the velocity in the
center-of-mass frame $\langle v(t)\rangle$, and time $t$, we 
can form two dimensionless quantities and write
\begin{equation}
\frac{\tau_{_{\!B}}}{t} = f\left(\frac{\langle \sigma\rangle/
\langle v\rangle}{t}\right)
\end{equation}
where $f$ is an unknown scaling function and the Boltzmann mean
collision time $\tau_{_{\!B}}(t)$ is the individual mean time 
between two collisions suffered by any one particle:
$\tau_{_{\!B}}$ is the ratio of the time-dependent mean-free-path 
to $\langle v\rangle$
\begin{equation}
\tau_{_{\!B}} = \frac{1}{n \langle \sigma \rangle \langle v \rangle}.
\label{eq:tau}
\end{equation}
$\langle \sigma\rangle/\langle v\rangle$ is the time needed for a typical
vortex to fly on the distance of a typical diameter, during which
the system suffers no collision. As a consequence of eq. (\ref{eq:sigma4}),
the surface coverage (or packing fraction) is indeed a decreasing
function of time, so that $\langle \sigma\rangle$ eventually becomes 
much smaller than the mean-free-path. Correspondingly, on a time 
scale $\langle \sigma\rangle/\langle v\rangle \ll \tau_{_{\!B}}(t)$, 
the evolution is free which indicates that $\tau_{_{\!B}}$ is the only
relevant time scale for the dynamics. Dimensional analysis then has to 
be reconsidered, and implies that $\tau_{_{\!B}}$ scales 
like the physical time
itself: $\tau_{_{\!B}}\propto t$. 

During a time interval $\tau_{_{\!B}}$, each vortex typically 
collides, leading to a decrease of density which is of the order 
of the density itself:
\begin{equation}
\frac{d\, n(t)}{dt} \propto - \frac{n(t)}{\tau_{_{\!B}}}.
\label{eq:varian}
\end{equation}
Inserting $\tau_{_{\!B}} \propto t$ into (\ref{eq:varian}), 
we obtain an algebraic 
time dependence for the density: this ``cri\-ti\-cal'' behaviour
can be attributed to the uniqueness of a relevant time scale in this
problem. 
We introduce two exponents $\xi$ and $\gamma$ to describe the asymptotic
regime: $n(t) \propto t^{-\xi}$, $\langle v(t)\rangle \propto t^\gamma$, 
the conservation of $\sigma^4$ during encounters implying
$n \langle \sigma^4\rangle \propto t^0$ and then 
$\langle \sigma\rangle \propto t^{\xi/4}$. 
From the definition (\ref{eq:tau}) and the relation $\tau_{_{\!B}} \propto t$, 
one obtains the corresponding scaling relation: 
\begin{equation}
\frac{3}{4}\, \xi  = 1 \,+\, \gamma.
\label{eq:scal}
\end{equation}
A more rigorous derivation, based on a scaling analysis
of the BBGKY-like hierarchy governing the evolution of the distribution 
functions can be found in \cite{TrHa}. We emphasize that eq. 
(\ref{eq:scal}) is exact in the scaling regime and 
independent of the coalescence rule 
defining the velocity of a vortex resulting from a colliding 
pair. 
An other but approximate relation between $\xi$ and $\gamma$
can be obtained from an analogy 
with a previously studied model of ballistic coalescence
\cite{CPY,TrHa}. On that purpose, 
we introduce for each vortex a fake mass 
$m_i \propto \sigma_i^4$ and define a fake energy by
$m_i v_i^2$. From eq. (\ref{eq:sigma4}), the mass is conserved 
during the collisions, which imposes that the average mass scales 
like the inverse of the density: $\langle m \rangle \propto t^\xi$. 
The total fake energy density $n \langle m v^2 \rangle$
consequently scales like $t^{-2\gamma}$. 
The second and approximate relation between $\xi$ and $\gamma$ is 
obtained assuming that each collision implies two typical uncorrelated 
particles, so that the fake energy loss equals the typical ({\it i.e.}
average) fake energy. The fake energy per particle is then conserved
(``isothermal'' approximation), and the fake energy density 
has the same temporal decay rate as the number density:
$\xi = -2\,\gamma$. Upon substitution into eq. (\ref{eq:scal}), 
one obtains 
\begin{equation}
\xi_{_{m\!f}}\, =\, \frac{4}{5}.
\label{eq:ximf}
\end{equation}
The above estimation may be coined ``mean-field'' and amounts 
to assuming that the linear momenta $m_i {\bf v}_i$
of the colliding vortices are uncorrelated \cite{These}.

Our estimate $\xi_{_{m\!f}}\! = 0.8$ is in fair 
agreement with the experimental 
and Navier-Stokes results, but needs to be confronted to numerical
simulations. On that purpose, we specify a transformation rule
for the velocity ${\bf v}_v$ of a vortex $v$ resulting form the merging
of vortices 1 and 2, by coming back to Kirchoff's law.
Equation (\ref{eq:kir}) can be recast in the form 
\begin{eqnarray}
{\bf v}_1 &=& \sum_{j \neq 1,2} \Gamma_j\, {\bf v}(j\to 1)\, +\, 
\Gamma_2\, {\bf v}(2\to 1) 
\label{eq:v1}\\
{\bf v}_2 &=& \sum_{j \neq 1,2} \Gamma_j\, {\bf v}(j\to 2) \,+\, 
\Gamma_1\, {\bf v}(1\to 2)\\
{\bf v}_v &=& \sum_{j \neq 1,2} \Gamma_j\, {\bf v}(j\to v)
\label{eq:v}
\end{eqnarray}
where ${\bf v}(j\to i)$ is the velocity field created at point
$i$ by a vortex $j$ of unit circulation. Just before the fusion, 
vortices 1 and 2 are close, and vortices $j\neq 1,2$ are on average 
far apart form the pair 1-2. We thus consider 
\begin{equation}
{\bf v}(j\to 1) \simeq {\bf v}(j\to 2) \simeq {\bf v}(j\to v) 
\quad \hbox{for } j\neq 1,2.
\label{eq:appr}
\end{equation}
Taking into account the symmetry constraint 
${\bf v}(j\to k) = - {\bf v}(k\to j)$, eqs. (\ref{eq:v1})-(\ref{eq:v})
supplemented with approximation (\ref{eq:appr}) yield 
\begin{equation}
\left(\Gamma_1  + \Gamma_2  \right) {\bf v}_v = 
\Gamma_1 {\bf v}_1 + \Gamma_2 {\bf v}_2 \qquad 
\Longleftrightarrow \qquad
{\bf v}_v = \frac{1}{\sigma_1^2+\sigma_2^2}\left(\sigma_1^2\,
{\bf v}_1 + \sigma_2^2 \, {\bf v}_2\right).
\label{eq:vnew}
\end{equation}
We have performed two-dimensional Molecular Dynamics (MD) simulations
of the ballistic coalescence model defined by merging rules 
(\ref{eq:sigma4}) and (\ref{eq:vnew}), with an initial number of vortices 
$5\,10^3 <N_0<5\,10^5$. Periodic boundary conditions
with a square simulation cell have been implemented and the
equilibrium state of elastic hard discs served as the initial condition. 
After an initial transient corresponding to $t<\tau_{_{\!B}}(t=0)$,
an algebraic scaling regime is evidenced, with a decay 
exponent $\xi\simeq 0.71 \pm 0.01$ (see figure 1)
independent of $N_0$ and of the initial packing fraction. 
The discrepancy between the mean-field prediction 
($\xi_{_{m\!f}}\!=0.8$) and the 
simulation result can be attributed to correlations in 
velocities required for the occurrence of a collision. These correlations
are neglected in mean-field, which overestimates exponent 
$\gamma$, and from (\ref{eq:scal}), overestimates $\xi$.
Our exponent is very close to the Navier-Stockes value $\xi \simeq 0.72$.
Moreover, the mean-free-path is found to scale like $t^{3\xi/4}$,
with a growth exponent 0.53, in agreement with the experimental
value 0.45$\,\pm\,$0.1 reported in \cite{HansenTabeling}.
These results  indicate that the essential features of the decay
stage can survive to a significant simplification of the complex
interactions between vortices. 

The robustness of $\xi$ with respect to the merging rule for velocity
has been tested by MD simulations with conservation of $\sigma^4$ 
and of $\sigma^4\bf{v}$ in collision events. This system is the
analog of the original ballistic coalescence model introduced
in \cite{CPY}, in the sense that the conserved linear momentum
$\sigma^4 \bf{v}$ involves a mass $\sigma^4$ which is itself
a conserved quantity (this is not the case with rules 
(\ref{eq:sigma4}) and (\ref{eq:vnew}) for which no conserved quantity
involving the velocity can be defined). The model again displays
an asymptotic density decay with $\xi \simeq 0.70 \pm 0.02$, irrespective
of the initial conditions. For the latter collision laws, the theory
of Smoluchowski rate equations \cite{vDEr} yields a mean-field approximation 
$\xi=4/5$ \cite{These}, confirming our estimate for $\xi_{_{m\!f}}$.

In summary, we have derived a ballistic coalescence model
mimicking the fusion of vortices in freely decaying two-dimensional 
turbulence. The decay exponent $\xi$ obtained is in agreement 
with recent laboratory experiments \cite{Marteau,HansenTabeling}
and close to its Navier-Stockes
counterpart. The precise form of the velocity merging
rule is found to have little influence on $\xi$ which depends
more crucially on the merging law for the size (eq. (\ref{eq:sigma4})).
Our simplification of the original punctuated-Hamiltonian
approach proposed by Carnevale {\it et al.} \cite{CmWPWY} further 
allows to put forward the simple mean-field analytical estimation
$\xi_{_{m\!f}}\!=0.8$, and 
indicates that conservation rule (\ref{eq:sigma4}) is the crucial
ingredient governing the self-similar stage of the decay.
However, 
the evolution does not conserve the surface coverage 
$\pi n \langle \sigma^2\rangle/4$, which decreases like $t^{-\xi/2}$:
the mean vortex size eventually becomes much smaller than the
typical distance $n^{-1/2}$ (itself vanishingly small with respect
to the mean-free-path), and a vortex can interact with 
many neighbours between two collisions in such a way that
its motion is likely to be quite erratic in the 
asymptotic regime. The validity 
of ballistic motion is thus questionable for very low
packing fractions, whereas it seems more reasonable for high
surface coverages ({\it i.e.} in the early stages of the dynamics).
The correspondence between the value of $\xi$ obtained here and
those reported previously \cite{CmWPWY,Weiss93,Marteau,HansenTabeling}
could then point out to the non asymptotic character of the latter. 
Finally, unlike the original turbulence solution \cite{Weiss93}, 
our model does not conserve the typical velocity 
$\langle v \rangle$ ($\gamma \neq 0$ and we measure $\gamma \simeq -0.47
\pm 0.01$): we argue that the relevant velocity for our 
model is the relative velocity of close vortices ({\it i.e.} 
the velocity at the scale of a typical diameter),
whereas the typical velocity of vortices associated with the 
conserved energy refers to the larger scale of 
inter-particle distances \cite{SommeriaP}. As we expect an energy
spectrum $E(k) \propto k^{-\alpha}$ with  $\alpha >1$, the ratio of these 
two quantities scales like
\begin{equation}
\frac{v(k=1/\langle \sigma \rangle)}{v(k=\sqrt n)}\, \propto \,
t^{\xi(1-\alpha)/8}
\end{equation}
and is thus a decreasing function of time. 

\vskip 2mm
{\em Acknowledgments.} The author acknowledges helpful discussions with
J.P. Hansen, P.H. Chavanis, J. Sommeria, P. Tabeling, A.E. Hansen,
T. Dombre and L. Bocquet.
The simulations were carried out under the auspices of the
P\^ole Scientifique de Mod\'elisation Num\'erique (Lyon). 
The hospitality of Professor Daan Frenkel 
(FOM Institute for Atomic and Molecular Physics, Amsterdam) 
is gratefully acknowledged.

\begin{figure}
\centerline{\hbox{
\psfig{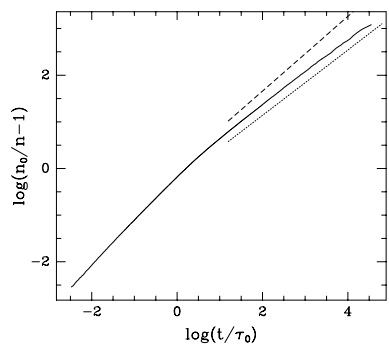}}}
\caption{ log$_{10}(n_0/n-1)$ vs. log$_{10}(t/\tau_0)$, where 
$\tau_0=\tau_{_{\!B}}(t\!=\!0)$ and
$n_0$ is the initial
surface density corresponding to a packing fraction 0.24. The initial
number of vortices is $N_0=10^5$. The dashed and dotted lines have
respectively slopes  
0.8 (mean-field prediction) and 0.71 }
\end{figure}

\end{document}

%% file: euromacr.tex
\newif\ifboo \boofalse
